\documentclass[runningheads]{llncs}
\usepackage[T1]{fontenc}
\usepackage{nicefrac}
\usepackage{subcaption}
\usepackage{hyperref}
\usepackage{graphicx}
\graphicspath{{images/}}
\usepackage{amsfonts}
\usepackage{color}

\usepackage{amsmath}
\usepackage{mathtools}
\newcommand{\orcid}[1]{\href{https://orcid.org/#1}{\includegraphics[width=10pt]{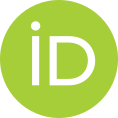}}}

\begin{document}
\title{Anisotropic Fanning Aware Low-Rank Tensor Approximation Based Tractography}
\titlerunning{Anisotropic Fanning Aware Low-Rank Tensor Tractography}
%
\author{Johannes Gruen\inst{1,2}\orcid{0000-0002-9154-3929} \and
	Jonah Sieg\inst{1}\orcid{0009-0002-0604-7320} \and
  Thomas Schultz\inst{2,1}\orcid{0000-0002-1200-7248}}
  \authorrunning{J. Gruen et al.}
%
  \institute{Institute for Computer Science, University of Bonn, Germany
   \and Bonn-Aachen International Center for Information Technology, Germany \\ 
   \texttt{schultz@cs.uni-bonn.de}}

\maketitle              
\begin{abstract}
  Low-rank higher-order tensor approximation has been used successfully to extract discrete directions for tractography from continuous fiber orientation density functions (fODFs). However, while it accounts for fiber crossings, it has so far ignored fanning, which has led to incomplete reconstructions. In this work, we integrate an anisotropic model of fanning based on the Bingham distribution into a recently proposed tractography method that performs low-rank approximation with an Unscented Kalman Filter. Our technical contributions include an initialization scheme for the new parameters, which is based on the Hessian of the low-rank approximation, pre-integration of the required convolution integrals to reduce the computational effort, and representation of the required 3D rotations with quaternions. Results on 12~subjects from the Human Connectome Project confirm that, in almost all considered tracts, our extended model significantly increases completeness of the reconstruction, while reducing excess, at acceptable additional computational cost. Its results are also more accurate than those from a simpler, isotropic fanning model that is based on Watson distributions.
\end{abstract}
\section{Introduction}
Diffusion MRI tractography \cite{Jeurissen:2018} permits the in-vivo
reconstruction of white matter tracts in surgery planning or scientific studies.
Spherical deconvolution is widely used to account for intra-voxel heterogeneity
by estimating a continuous fiber orientation density function (fODF) in each
voxel \cite{DellAcqua:2018}. Representing fODFs as higher-order tensors and
applying a low-rank approximation to these tensors has been shown to be a robust and
efficient approach to estimating discrete tracking directions
\cite{lowrank,Ankele:CARS2017}.

However, while low-rank approximation accounts for fiber crossings, it ignores fiber fanning \cite{Sotiropoulos:2012}. Consequently, even though recent work \cite{Gruen:2023} has achieved promising results by performing low-rank approximation within the framework of Unscented Kalman Filter (UKF) based tractography \cite{Malcolm:MedIA2010}, some fanning bundles were extracted incompletely when using single-region seeding strategies \cite{Gruen:2023}.

We address this limitation by explicitly modeling anisotropic fanning in the
low-rank UKF with Bingham distributions \cite{Kaden:2007}. This involves three
main technical challenges: Firstly, initializing additional parameters in the
UKF state. Section~\ref{sec:hessian-initialization} solves this by observing
that the Hessian matrix at the optimum of the low-rank approximation indicates
the amount and direction of fanning. Secondly, the computational effort of
convolving rank-one tensors with Bingham distributions.
Section~\ref{sec:convolution} solves this by pre-computing the corresponding
integrals and storing results in lookup tables. Thirdly, maintaining a full 3D rotation per fiber compartment. Section~\ref{sec:quaternions} solves this with a quaternion-based representation. Results in Section~\ref{sec:results} indicate that our extension reconstructs fanning bundles significantly more completely, while reducing excess, at acceptable additional computational cost.

\section{Background and Related Work}
\subsection{Low-rank tensor approximation model}
Constrained spherical deconvolution (CSD) computes the
fiber orientation distribution function (fODF), a mapping from
the sphere to $\mathbb{R}_+$ which captures the
fraction of fibers in any direction \cite{TOURNIER20071459}. One widely used strategy for estimating principal fiber orientations is to consider local fODF maxima. Our work builds on a variation of CSD, which
represents the fODF as a symmetric higher-order tensor $\mathcal{T}$ and estimates $r$ fiber
directions via a rank-$r$ approximation
\begin{align}
	\mathcal{T}^{\left( r \right)} = \sum_{i=1}^r \alpha_i
	\mathbf{v}_i^{\otimes l}, 
	\label{eq:low-rank}
\end{align}
where the scalar $\alpha_i \in \mathbb{R}_+$ denotes the volume fraction of the
$i$th fiber, $\mathbf{v}_i \in \mathbb{S}^2$ its direction, and the superscript $\otimes l$ indicates an $l$-fold symmetric outer product, which turns the vector into an order-$l$ tensor. The main benefit of this approach is that it can separate crossing fibers even if they are not distinct local maxima, which permits the use of lower orders and in turn improves numerical conditioning and computational effort \cite{Ankele:CARS2017}. Specifically, the angular resolution of fourth-order tensor approximation for crossing fibers has been shown to exceed order-eight fODFs with peak extraction \cite{Ankele:CARS2017}. To additionally capture information about anisotropic fanning, our current work increases the tensor order to $l=6$, which parameterizes each fODF with 28 degrees of freedom.

\subsection{Bingham distribution}

The Bingham distribution \cite{bingham} is the spherical and antipodally symmetric
$(f\left( \mathbf{x} \right) = f \left( -\mathbf{x} \right))$ analogue
to a two dimensional Gaussian distribution. It is given by the probability density
function
\begin{align}
	f \left( \mathbf{x}; \mathbf{M}, \mathbf{Z} \right) \coloneqq \frac{1}{N \left(
	\mathbf{Z} \right)} \exp \left( \mathbf{x}^T \mathbf{M} \mathbf{Z}
		\mathbf{M}^T \mathbf{x}
	\right),
	\label{eq:bingham}
\end{align}
where $\mathbf{Z}$ is a diagonal matrix with decreasing entries $z_1 \geq
z_2 \geq z_3$, $\mathbf{M} = \left( \mu_1, \mu_2, \mu_3 \right)$ is an orthogonal matrix and 
 $N \left( \mathbf{Z} \right)$ denotes the hypergeometric function  of matrix argument. Without loss of generality, we set
$z_3 = 0$ and rename $\kappa = z_1, \beta = z_2$ to rewrite the
density function as:
\begin{align}
	f \left( \mathbf{x}; \mu_1, \mu_2, \kappa, \beta \right) =  \frac{1}{N \left(
	\kappa, \beta \right)} \exp \left( \kappa \langle \mu_1, \mathbf{x} \rangle^2
	+ \beta \langle \mu_2, \mathbf{x} \rangle^2\right).
\end{align}

The Bingham distribution was used previously to model anisotropic fanning: Riffert et al.\ \cite{RIFFERT} fitted a mixture of Bingham distributions to the fODF to compute metrics such as peak spread and integral over peak. Kaden et al. \cite{Kaden:2007} used it for Bayesian tractography. Our contribution combines the Bingham distribution with the low-rank model, and estimates the resulting parameters with a computationally efficient Unscented Kalman Filter.

\subsection{Unscented Kalman Filter}
The Kalman Filter is an algorithm that estimates a set of unknown variables,
typically referered to as the state, from
a series of noisy observations over time. The Unscented Kalman Filter (UKF)
\cite{UKF} is an extension that permits a non-linear relationship between the
unknown variables and the measurements. It has first been used for tractography
by Malcolm et al. \cite{filteredMultiTensor,Malcolm:MedIA2010}, who treat the
diffusion MR signal as consecutive measurements along a fiber, and the parameters of a mixture of diffusion tensors \cite{filteredMultiTensor} or Watson distributions \cite{Malcolm:MedIA2010} as the unknown variables. Compared to independent estimation of model parameters at each location, this approach reduces the effects of measurement noise by combining local information with the history of previously encountered values. Consequently, it has been used for scientific studies \cite{Chen:2016,Dalamagkas} as well as neurosurgical planning \cite{Cheng:2015}.

Recent work has used the UKF to estimate the parameters of the low-rank model \cite{Gruen:2023}. This variant of the UKF treats the fODFs instead of the raw diffusion MR signal as its measurements, which increases tracking accuracy while reducing computational cost, due to the much lower number of fODF parameters compared to diffusion-weighted volumes. A remaining limitation of that approach is that it does not account for fanning.

\section{Material and Methods}

We extend the previously described low-rank UKF \cite{Gruen:2023} by modeling directional fanning with a Bingham distribution (Section~\ref{sec:low-rank-bingham}). Implementing this requires solving problems related to initialization (Section~\ref{sec:hessian-initialization}), efficient evaluation of certain integrals (Section~\ref{sec:convolution}), and representing rigid body orientations within the UKF (Section~\ref{sec:quaternions}). Section~\ref{sec:tractography} describes the resulting tractography algorithm, while Section~\ref{sec:data} reports the data and measures that we use for evaluation.

\subsection{Low-rank model with anisotropic fanning}
\label{sec:low-rank-bingham}
The higher-order tensor variant of CSD adapts the deconvolution so that it maps the single fiber response to a rank-one tensor \cite{lowrank}. Therefore, fanning can be incorporated by convolving the rank-$1$ kernel $k$ with the Bingham distribution
\begin{align}
	h^{\left( r \right)} = \sum_{i = 1}^{r} \alpha_i f \left( \cdot;
		\mu^{\left( i \right)}_1,
		\mu^{\left( i \right)}_2, \kappa^{\left( i \right)},
		\beta^{\left( i \right)}
	\right) \star k ,
	\label{eq:bingham-convolution}
\end{align}
where $\alpha_i$ denotes the volume fraction of the $i$th fiber in direction
$\mu^{\left( i \right)}_1$, $\kappa^{(i)}$ the concentration around it (i.e.,
the inverse to the amount of fanning). In case of anisotropic fanning,
$\beta^{(i)}>0$ indicates the additional amount of fanning in direction
$\mu^{\left( i \right)}_2$. For $\kappa^{(i)} \rightarrow \infty$ and
$\beta^{(i)} = 0$, the Bingham distributions converge to delta peaks and
the model (\ref{eq:bingham-convolution}) converges towards the original low-rank model
(\ref{eq:low-rank}) with fiber directions $\mu^{\left( i \right)}_1=\mathbf{v}_i$.

\subsection{Initialization via the low-rank model}
\label{sec:hessian-initialization}

Since it is difficult to fit the model in Eq.~(\ref{eq:bingham-convolution}) to
data, we initialize the UKF based on the original low-rank approximation in
Eq.~(\ref{eq:low-rank}). Firstly, we use the same main fiber directions,
$\mu^{\left( i \right)}_1=\mathbf{v}_i$. Secondly, we initialize the fanning related parameters by observing that the rate at which the approximation error grows when rotating a given fiber direction away from its optimum depends on the amount of fanning: The lower the amount of fanning (the sharper the fODF peak), the more sensitive is the approximation error to the exact direction.

For each fiber, this information is captured in the second derivatives of the cost function with respect to its orientation, i.e., a $2\times 2$ Hessian that can be computed in spherical coordinates; an equation for this is derived in \cite{schultz:hessian}. There is a one-to-one mapping between the eigenvalues of that Hessian and corresponding values of $\kappa$ and $\beta$. The eigenvector corresponding to the lower eigenvalue indicates the dominant fanning direction $\mu_2$.

\begin{figure}[tb]
	\begin{subfigure}[b]{0.49\textwidth}
		\centering
		\includegraphics[width=\textwidth]{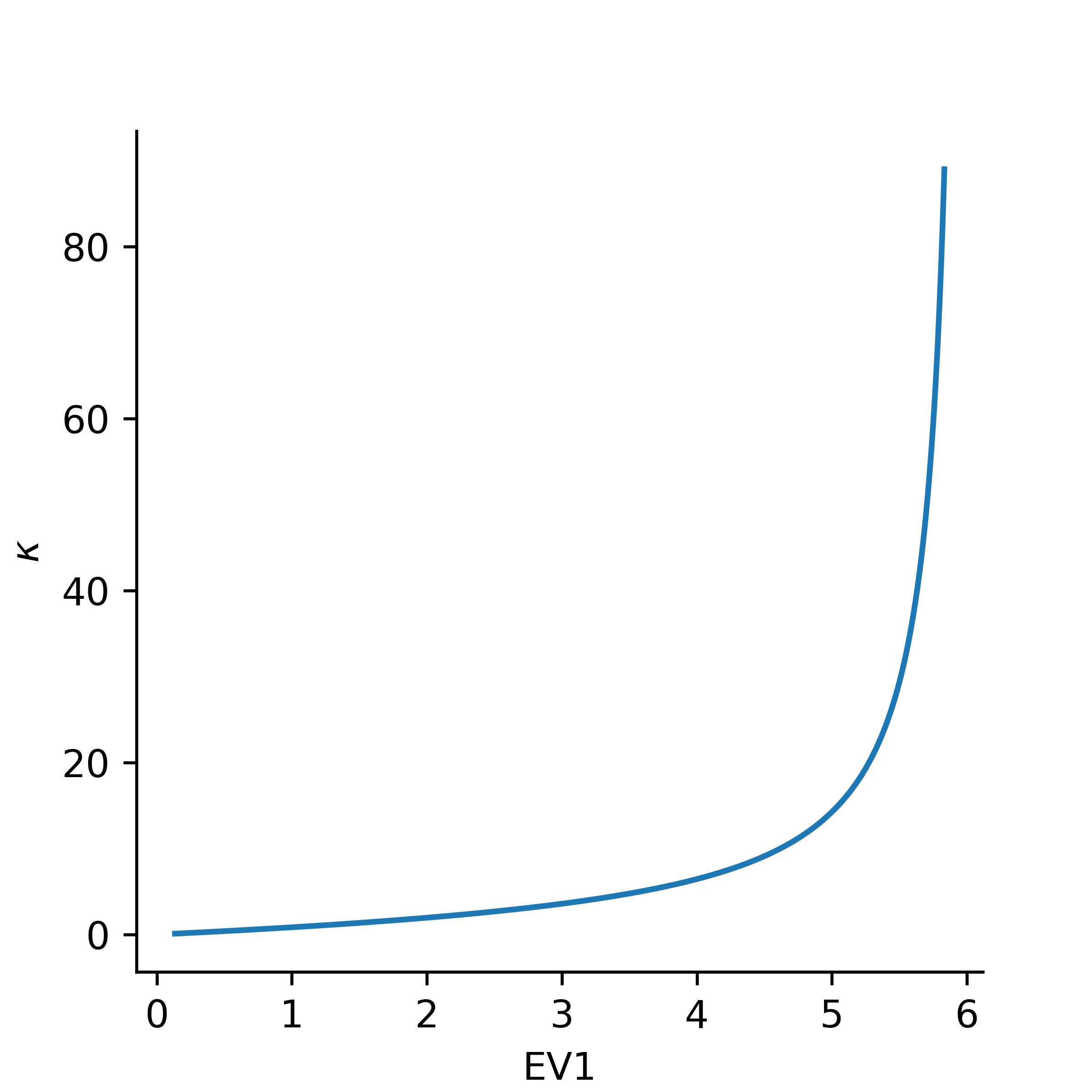}
		\caption{Mapping from larger eigenvalue to $\kappa$}
		\label{fig:mapping_ew_kb:a}
	\end{subfigure}
	\begin{subfigure}[b]{0.49\textwidth}
		\centering
		\includegraphics[width=\textwidth]{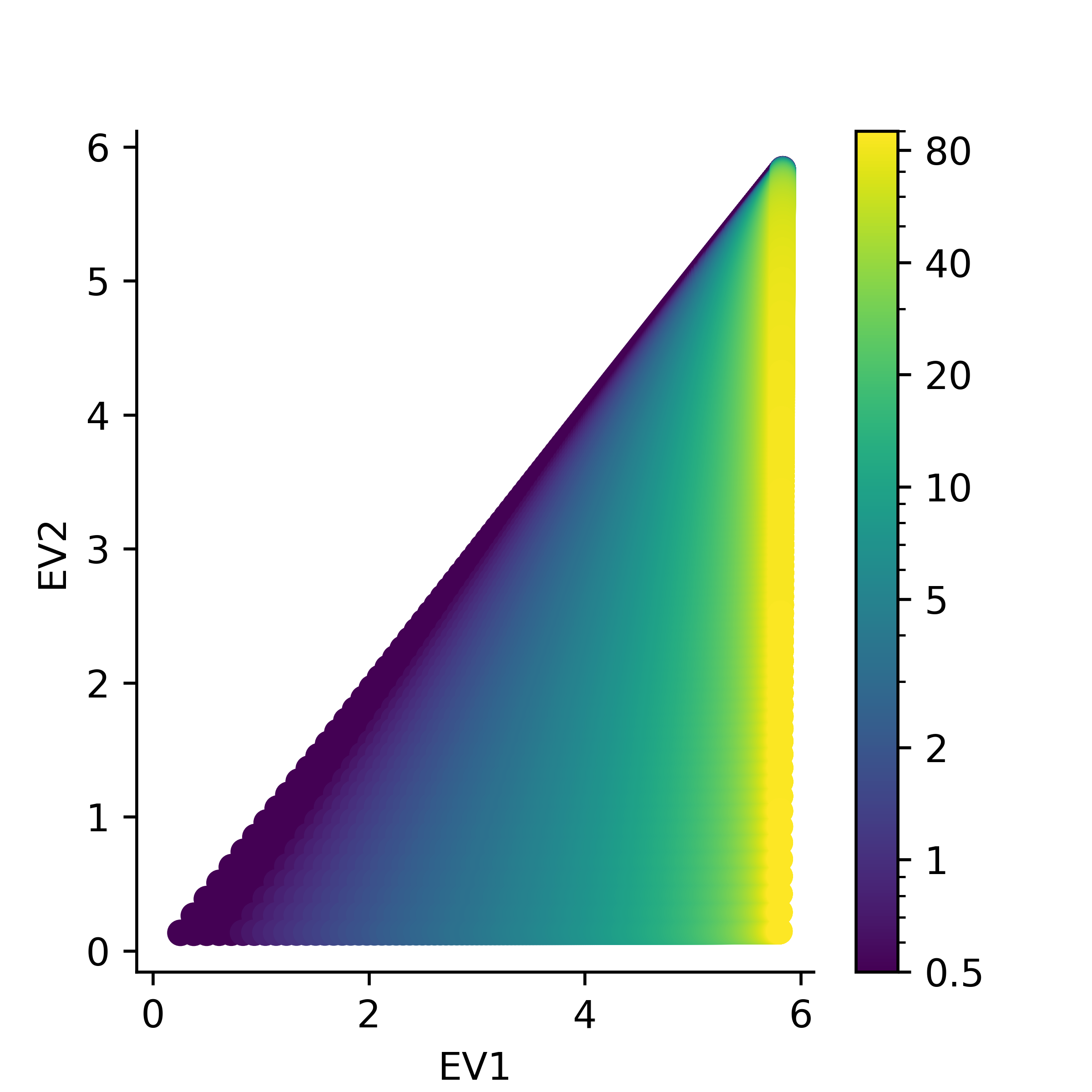}
		\caption{Mapping from eigenvalues to $\beta$}
		\label{fig:mapping_ew_kb:b}
	\end{subfigure}
	\caption{Lookup tables for initializing $\kappa$ and
	$\beta$ based on Hessian eigenvalues. Left: $\kappa$ is fully determined by the larger 
	eigenvalue. Right: $\beta=0$ if both eigenvalues are the same, but increases as EV2 decreases for a fixed value of EV1.}
	\label{fig:mapping_ew_kb}
\end{figure}

We pre-compute a lookup table for the values of $\kappa$ and $\beta$, given the Hessian eigenvalues. To this end, we utilize the model (\ref{eq:bingham-convolution}) to generate single fiber
fODFs for various combinations of
$\kappa$ and $\beta$ values, and record the resulting eigenvalues. Figure \ref{fig:mapping_ew_kb} visualizes the mapping from eigenvalues to $\kappa$ and $\beta$. Subfigure \ref{fig:mapping_ew_kb:a} shows that $\kappa$ increases with the larger 
eigenvalue, indicating a higher concentration around the main fiber direction. Subfigure \ref{fig:mapping_ew_kb:b} shows how $\beta$ depends on both eigenvalues. If they are the same, fanning is isotropic ($\beta=0$), while for any given larger eigenvalue (EV1), $\beta$ increases, indicating an increasingly elliptic fanning, as the smaller eigenvalues (EV2) decreases towards zero.

We apply this lookup table to multi-fiber voxels by computing the residual fODF for each fiber (i.e., we subtract out the remaining fibers), and normalizing it such that $\alpha=1$ to eliminate scaling effects. After fixing all fiber directions and fanning parameters, we fit the remaining volume fractions $\alpha_i$ in Eq.~(\ref{eq:bingham-convolution}) with a non-negative least squares solver. 

\subsection{Pre-computing the convolution}
\label{sec:convolution}
Equation~(\ref{eq:bingham-convolution}) involves a
 convolution between a rank-$1$ kernel and
 a Bingham distribution. To compute it efficiently, we first split the Bingham distribution into a standard version and a rotation part. 
We rewrite
\begin{align}
	f \left(
	\mathbf{x};  \mathbf{M}, \mathbf{Z} \right) = \left( D \left( \vartheta,
			\psi, \omega
	\right) g \right) \left( \mathbf{x}; \kappa, \beta \right) = g \left(
		\mathbf{M}^{-1} \mathbf{x}; \kappa, \beta
	\right),
\end{align}
where $g \left( \mathbf{x}, \kappa, \beta \right) \coloneqq \frac{1}{N \left(
		\mathbf{Z}
\right)} \exp \left( \kappa \mathbf{x}_3^2 + \beta \mathbf{x}_2^2 \right)$ is a
standard Bingham distribution in the canonical basis oriented towards the north pole and
$D$ is the $zyz$ rotation matrix, which is defined as
\begin{align}
	\mathbf{M} = D \left( \vartheta, \psi, \omega \right) \coloneqq R_z \left( \vartheta
	\right) R_y \left( \psi \right) R_z \left( \omega \right)
\end{align}
with 
\begin{align}
	R_z \left( \alpha  \right) \coloneqq \left( 
	\begin{matrix}
		\cos \alpha & \sin \alpha & 0 \\
		- \sin \alpha & \cos \alpha & 0 \\
		0 & 0 & 1 \\
\end{matrix} \right)
\text{ and } 
	R_y \left( \alpha  \right) \coloneqq \left( 
	\begin{matrix}
		\cos \alpha & 0 & \sin \alpha \\
		0 & 1 & 0 \\
		- \sin \alpha & 0 & \cos \alpha  \\
\end{matrix} \right) .
\end{align}

This decomposition 
is a significant simplification, because
we can now pre-compute the convolution between the standard Bingham
distribution and the kernel, and apply the rotation afterwards.

As it is standard practice in CSD \cite{TOURNIER20071459}, we perform the convolution on the sphere using spherical and rotational harmonics. A rotational harmonics representation of the rank-1 kernel has been computed previously \cite{lowrank}. Unfortunately, no closed form solution is available for the spherical harmonics coefficients of the Bingham distribution. Therefore, we pre-compute them numerically, for the relevant range of $\kappa \in \left\{ 2.1, 2.2 , \dots , 89 \right\}$ and
$\beta \in \left\{ 0, 0.1, \dots , \kappa - 2  \right\}$.

\subsection{Representing rotations with quaternions}
\label{sec:quaternions}
Unlike previous UKF-based tractography methods, our model requires a full three-dimensional rotation per fiber to account not just for the fiber direction, but also for the direction of its anisotropic spread. Unit quaternions are a popular representation of rotations, since they overcome limitations of Euler angles, such as gimbal lock. However, integrating them into a UKF is non trivial, since their normalization leads to
dependencies within the state \cite{Kraft}. We
overcome the problem by utilizing a homeomorphism between quaternions and $\mathbb{R}^3$. We discuss the relevant steps of that approach, but refer the reader to work by Bernal-Polo et al.\ \cite{MUKF} for a more detailed discussion of quaternions and this way of integrating them into the UKF. More detailed explanations of UKF-based tractography are also available in the literature \cite{filteredMultiTensor,Malcolm:MedIA2010,Gruen:2023}.

Given a quaternion $q = \left[ q_w, q_x, q_y, q_z \right] \in \mathbb{H}$,
we define a homeomorphism
\begin{align}
	\phi :  \mathbb{S}^3 & \rightarrow
	\left\{ \mathbf{e} \in \mathbb{R}^3 : \| \mathbf{e} \| \leq 4 \right\} \\
		q & \mapsto  4 \frac{q_{1:}}{1+q_0}
\end{align}
to the so-called Modified Rodrigues Parameters \cite{MRP} and, vice versa,
\begin{align}
	\phi^{-1} : \left\{ \mathbf{e} \in \mathbb{R}^3 : \| \mathbf{e} \| \leq 4 \right\} 
	 & \rightarrow   \mathbb{S}^3   \\
	 \mathbf{e} & \mapsto  \frac{1}{16+\| \mathbf{e} \|^2} \left( 16 - \|
	 \mathbf{e} \|^2, 8\mathbf{e} \right).
\end{align}

In a close neighborhood of
the identity quaternion, these charts behave like the identity transformation
between the imaginary part of quaternions and $\mathbb{R}^3$. For a  given mean
quaternion $\bar{q}$ of a quaternions set $ \left\{ q_i \right\}_i$, we define a mapping which
first maps each $q_i$ by the conjugated mean quaternion, pushes it to
$\mathbb{R}^3$,
\begin{align}
	\phi_{\bar{q}}\left( q_i \right) \coloneqq \phi \left( \bar{q}^{\star} \star q_i \right) =\mathbf{e},
\end{align}
where $\bar{q}^{\star} = \left[ \bar{q}_w, -\bar{q}_x, -\bar{q}_y, -\bar{q}_z
\right]$ denotes the conjugated quaternion, pulls it back and rotates it back via
\begin{align}
	\phi^{-1}_{\bar{q}} \left( \mathbf{e} \right) \coloneqq \bar{q} \star \phi^{-1}
	\left( \mathbf{e} \right) . 
\end{align}
Assuming that the quaternions are highly concentrated around the mean
quaternion, the embedding resembles the distribution of quaternions closely.
\begin{figure}[h]
	\centering
	\includegraphics[width=\linewidth]{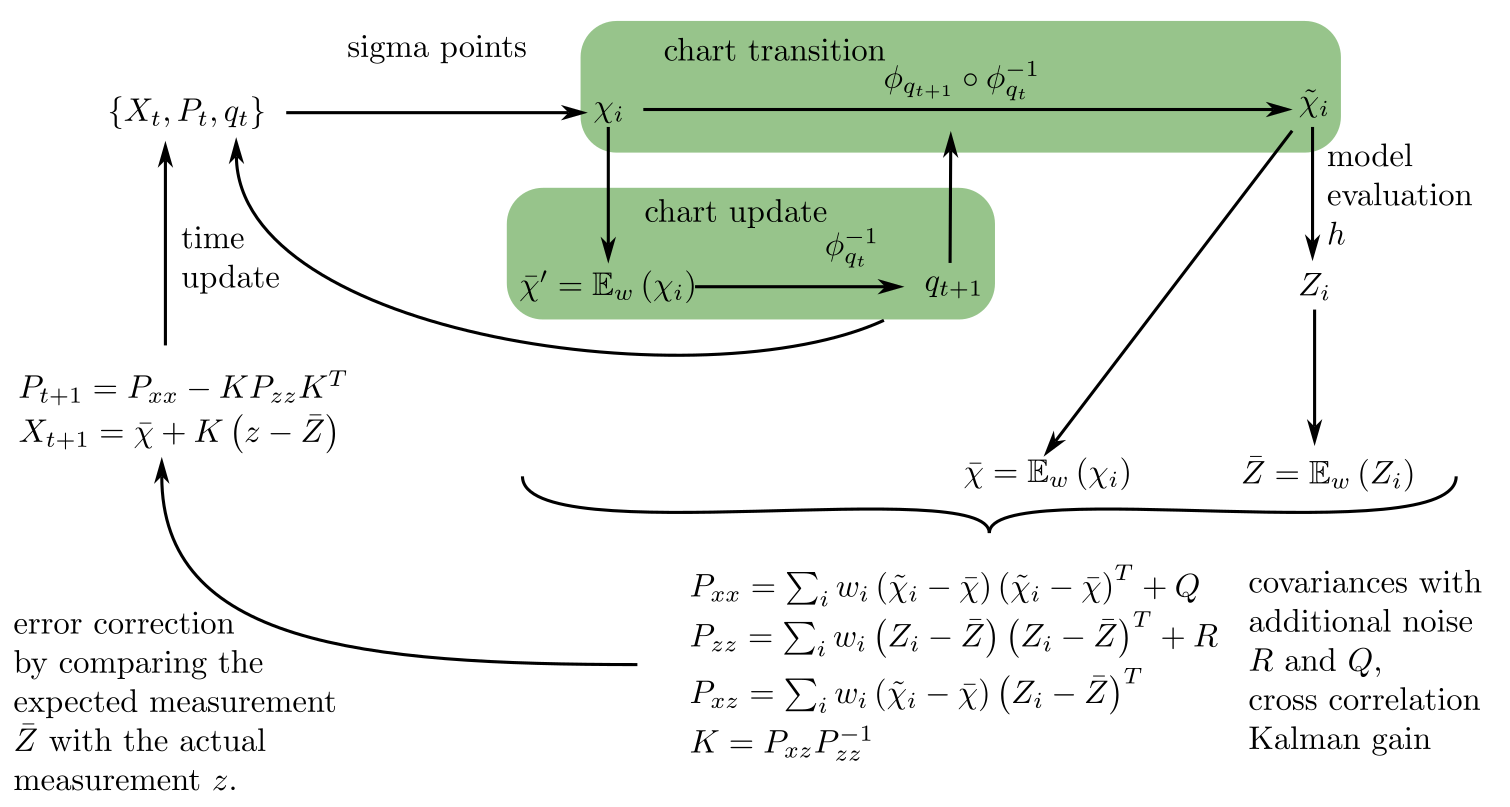}
	\caption{Schematic representation of an UKF update step. Compared to a
		normal UKF update, we additionally update the charts by first
		calculating the weighted mean of the sigma points and pulling it
		back into quaternion space. With the new quaternion, we perform
		a chart transition. The remaining steps are analogue to the
	traditional UKF.}
	\label{fig:ukf}
\end{figure}

With these preliminaries, we set up the UKF as illustrated in Figure \ref{fig:ukf}. We only show it for a single
fiber direction. For simplicity, our implementation updates the parameters for each fiber separately. The state at point $t$ is
defined by the parameters of a single Bingham distribution in the embedded space 
\begin{align}
	X_t \coloneqq \left\{ \alpha, \kappa, \beta, \mathbf{e}_1, \mathbf{e}_2,
		\mathbf{e}_3
	\right\}. 
\end{align}
The embedding is fully determined by the quaternion $q_t$ and the covariance is
denoted by $P_t$.

We create sigma points to capture the distribution of the covariance around the
current mean. We use the sigma points to calculate a chart update $q_{t+1}$, by taking a weighted mean with weights $w_i$ and pulling the embedded part back into quaternion space. With the new chart we
perform a chart transition. Afterwards, we follow the standard UKF update
scheme: Firstly, calculate the weighted mean of the sigma points, 
evaluate our model for all sigma points and take the corresponding weighted mean.
Secondly, calculate the covariance $P_{xx}$ of the sigma points, the
covariance of the evaluation $P_{zz}$ and the cross correlation $P_{xy}$. This
information is then used to calculate the Kalman gain $K$ and correct the
current state $X_t$ dependent on the difference between the expected measurement
$\bar{Z}$ and the fODF $z$ as well as the covariance.
\subsection{Probabilistic streamline-based tractography}
\label{sec:tractography}

For a given seed point, we initialize the UKF as discussed in Section~\ref{sec:hessian-initialization}. We perform streamline integration with second-order Runge-Kutta: At the $j$th point of the streamline, we 
update the UKF, select the Bingham
distribution whose main direction is closest by angle to the current tracking direction, and draw a direction from that Bingham distribution via rejection sampling. We use that direction for a tentative half-step, again update the UKF and perform rejection sampling. Finally, we reach point $(j+1)$ by taking a full step from point $j$ in that new direction. This process is iteratively conducted
until a stopping criterion is reached. We stop the integration if the white matter density drops below $0.4$ or if we
cannot find any valid direction within 60 degrees.

\subsection{Data and evaluation}
\label{sec:data}
It is the goal of our work to modify the UKF so that it more completely reconstructs fanning bundles from seeds in a single region. We evaluate this on 12~subjects from the Human Connectome
Project (HCP) \cite{HCP} for which
reference tractographies have been published as part of
TractSeg \cite{WASSERTHAL2018239}. They are based on a segmented and manually refined whole-brain tractography. We evaluate reconstructions of these tracts from seed points that we obtain by intersecting the reference
bundles with a plane, and picking the initial tracking direction that is closest to the reference fiber's tangent at the seed point. We estimate fODFs using data from all three $b$ shells that are available in the HCP data \cite{Ankele:CARS2017}.

We use a step size of $0.5$~mm and seed 3 times at
each seed point. Due to the probabilistic nature of our method, we perform density filtering to remove single outliers. Since diffusion MRI tractography is known to create false positive streamlines \cite{ismrm}, we also apply filtering based on inclusion and exclusion regions similar to the ones described by Wakana et al. \cite{Wakana:2007}. We place those regions manually in a single subject, and transfer them to the remaining ones via linear registration. Any streamline that does not intersect with all inclusion
regions or intersects with an exclusion region is removed entirely. 

To make the comparison against the previously described low-rank UKF \cite{Gruen:2023} more direct, we set its tensor order to $6$. We also evaluate the benefit of modeling \emph{anisotropic} fanning by implementing a variant of our approach that uses an isotropic Watson distribution, and could be seen as an extension of the previously proposed Watson UKF \cite{Malcolm:MedIA2010}. For this
model and for the Bingham UKF, we conducted a grid search to tune parameters and
finalized $Q = \left\{ \alpha =0.05, \kappa =0.05, v_1 = 0.02, v_2 =
0.02,v_3= 0.02 \right\}$ and $R=0.02$ for the Watson UKF and $Q = \left\{
	\alpha = 0.01, \kappa=0.1, \beta=0.1, e_1=0.005, e_2=0.005, e_3=0.005
\right\}$ and $R=0.02$  for the Bingham UKF. For all models we set the fiber
rank to 2.

We judge the completeness and excess of all tractographies based on distances between points on the reference tracts, and the generated ones. Specifically, we employ the $95\%$ quantile $\chi^{95\%}$ of the  directed Hausdorff distance 
\begin{align}
	h \left( A,B \right) \coloneqq \chi^{95\%} \left\{ \min_{\mathbf{b} \in B} \|
	\mathbf{a} - \mathbf{b} \| : \mathbf{a} \in A \right\} ,
	\label{eq:directed-hausdorff}
\end{align}
where $A$ and  $B$ denote point sets \cite{directedHausdorff}. Intuitively, if the $95\%$ quantile
of $h(A, B) = d$, then $95\%$ of the vertices of $A$ are within distance $d$
from some point of $B$. This measure is not symmetric.
Thus, setting $A$ to the reference tractography and $B$
to the reconstruction penalizes false negatives (it scores completeness), while switching the arguments penalizes false positives (it scores the excess).

\section{Results}
\label{sec:results}
\begin{figure}[t]
	\centering
	\includegraphics[width=\linewidth]{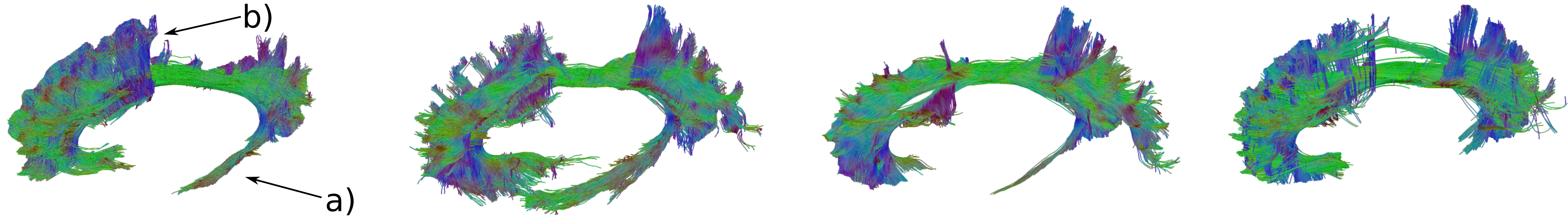}
	\caption{Reconstructions of the left Cingulum. From left to
	right: Reference, Bingham, Watson, low-rank UKF. The
	Bingham UKF permits the most complete reconstruction. The
low-rank UKF misses the  parahippocampal part a), while the Watson UKF misses fibers towards the anterior cingulate cortex
b).}
	\label{fig:CG}
\end{figure}

\begin{figure}[b]
	\centering
	\includegraphics[width=\linewidth]{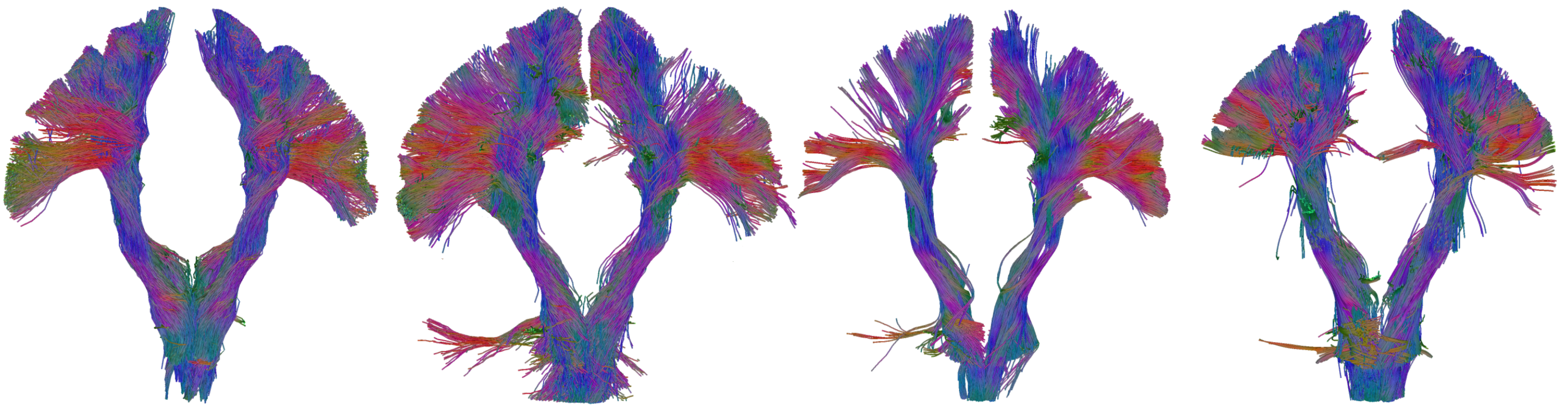}
	\caption{Reconstructions of the corticospinal tract (CST). From left to
	right: Reference, Bingham, Watson, low-rank UKF. The Bingham UKF
leads to the highest streamline density in the lateral fanning.}
	\label{fig:CST}
\end{figure}

Figure \ref{fig:CG} presents a qualitative comparison of the reconstruction of
the Cingulum (CG) in an example subject. In comparison to the low-rank UKF, both the Watson
UKF and the Bingham UKF result in a more complete reconstruction of the
parahippocampal part a). Moreover, compared to the Watson UKF, the Bingham UKF
achieves a more complete reconstruction of fibers entering the anterior cingulate cortex b). Similar trends are observed in Figure \ref{fig:CST} for the
reconstruction of the cortospinal tract (CST). The
Bingham UKF successfully reconstructs a majority of the lateral fibers, while
both the low-rank UKF and the Watson UKF are missing some parts of the fanning.

\begin{figure}[t!]
	\centering
	\includegraphics[width=\linewidth]{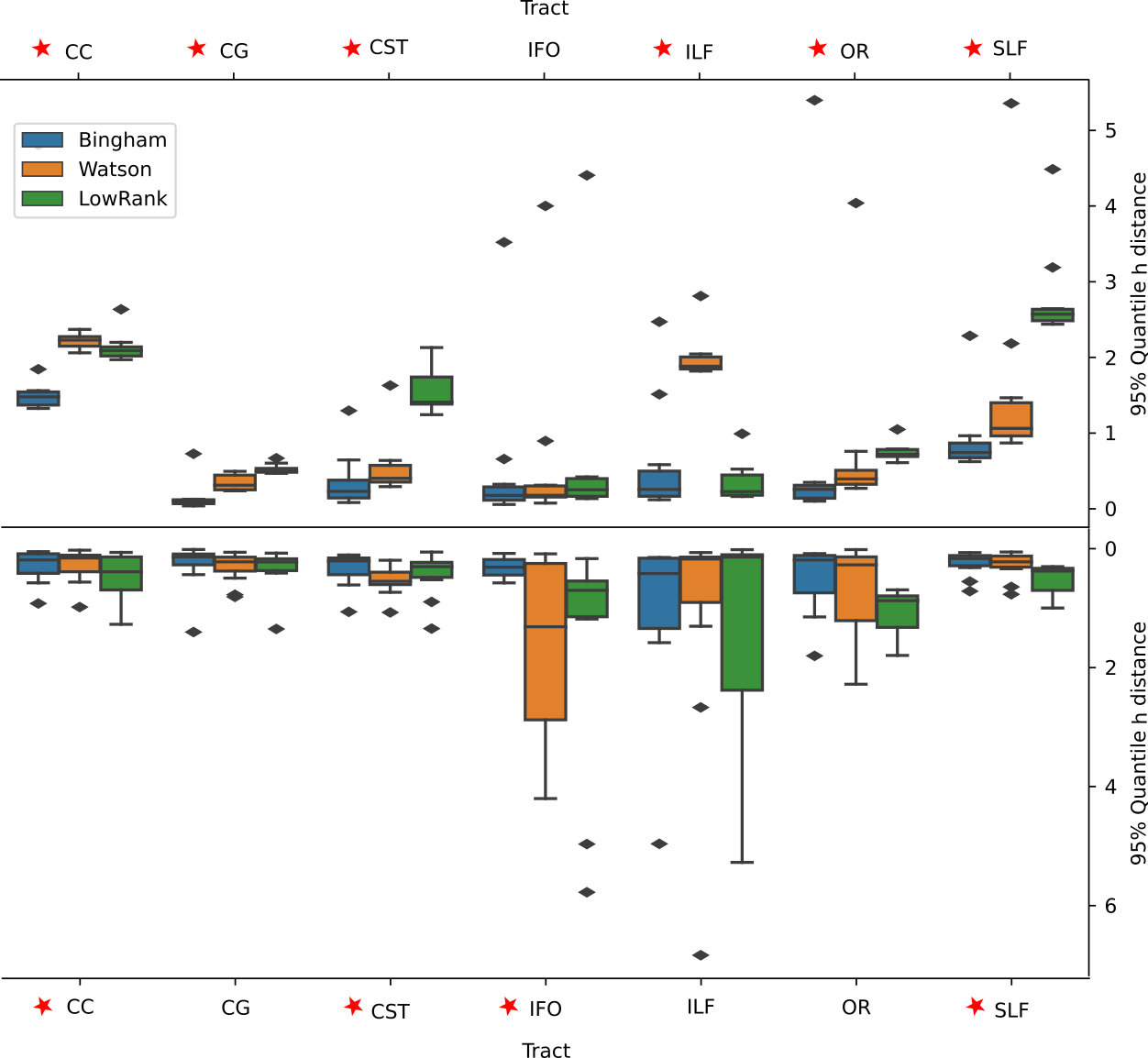}
	\caption{Top: 95\% quantile of the directed Hausdorff distance from reference
          to reconstruction. Median distances are lowest, indicating highest completeness, for Bingham UKF in all tracts except the ILF. In 6 out of 7 tracts, differences are significant (asterisks).
          Bottom: Directed distance from reconstruction to reference.          In most tracts, even these distances decrease when modeling fanning, indicating improved specificity in addition to the higher completeness.}
	\label{fig:hd-measure}
\end{figure}

We quantify these results by evaluating directed Hausdorff
distances. The upper part of Figure \ref{fig:hd-measure} shows
distances from the reference to the reconstruction. In 6 out of 7 tracts, the Bingham UKF exhibits the lowest median, indicating the most complete reconstructions.
The lower part measures distances from the reconstruction to the reference, so that low values indicate low excess. In 6 out of 7 tracts, the Bingham UKF leads to a lower median than the low-rank UKF, indicating that specificity is improved in addition to the increased sensitivity.

To statistically assess the differences between the proposed methods, we
conducted a Friedman test  \cite{doi:10.1080/01621459.1937.10503522} for each tract. An asterisk denotes
significant differences at
significance level of $p < 0.007$, due to Bonferroni correction. In 6 out of 7
tracts, we found significant differences in the completeness of reconstruction.
In 4 out of 7 tracts, significant differences were observed for the
excess. 

Generation of 1000 CST streamlines took 92.5 seconds for the Bingham
UKF, 85.2 seconds for the Watson UKF, and 58.1 seconds for the low-rank UKF on a
single core of a 3.3 GHz CPU.  
\section{Conclusion}
We developed a new algorithm for probabilistic tractography that incorporates anisotropic fanning into the recently described
low-rank UKF. We demonstrated that this
results in more complete reconstructions, while also reducing false positives, in almost all bundles. Our proposed technical solutions for initialization, convolution, and representation of rotations contribute to maintaining acceptable computational efficiency. Our code will be made available along with the publication.

\section*{Acknowledgment}
Funded by the Deutsche Forschungsgemeinschaft (DFG, German Research Foundation) – 422414649. Data were provided by the
Human Connectome Project, WU-Minn Consortium (Principal Investigators: David Van Essen and Kamil Ugurbil; 1U54MH091657) funded
by the 16 NIH Institutes and Centers that support the NIH Blueprint for Neuroscience Research; and by the McDonnell Center for Systems
Neuroscience at Washington University

\bibliographystyle{splncs04}
\bibliography{sn-bibliography}
\end{document}